\documentclass{PoS}
\pdfoutput=1

\title{Measurement of the Longitudinal Single-Spin Asymmetry for $W$ Boson Production at STAR}

\ShortTitle{Measurement of $A_L$ for $W$ Boson Production at STAR}

\author{\speaker{Jinlong Zhang, for the STAR Collaboration}%
       \\ Lawrence Berkeley National Laboratory 
       \\ Shandong University
       \\E-mail: \email{jlzhang@lbl.gov}}

\abstract{The production of $W^\pm$ bosons in longitudinally polarized $p+p$ collisions at RHIC  provides a direct probe for the spin-flavor structure of the proton through the parity-violating single-spin asymmetry, $A_L$. At STAR, the leptonic decay channel $W \to e\nu$ can be measured with the electromagnetic calorimeters and time projection chamber. STAR has previously measured $A_L$ as a function of the decay electron and decay positron pseudorapidities from datasets taken in 2011 and 2012. This has provided significant constraints on the $\bar{u}$ and $\bar{d}$ quark helicity distributions. 
In 2013 the STAR experiment collected an integrated luminosity of $\sim$300 pb$^{-1}$ at $\sqrt{s}=510$ GeV with an average beam polarization of $\sim$56\%, which is  more than three times larger than the total integrated luminosity of previous years. The new preliminary results of the $W$ $A_L$ analysis for the dataset collected in 2013 are reported. }

\FullConference{The 26th International Nuclear Physics Conference\\
         11-16 September, 2016\\
         Adelaide, Australia}

\begin{document}

\section{Introduction}

The proton's spin structure has attracted theoretical and experimental interest in the past few decades. Polarized inclusive deep-inelastic scattering (DIS) experiments have provided data from which it was found that the quark and antiquark spins only contribute $\sim$30\% of the proton spin~\cite{sumR}. In semi-inclusive DIS measurements, the flavor decomposition of quark spin contribution to proton spin can be accessed by identifying one or more hadrons in final state. Fragmentation functions are required to relate the final state hadrons to the scatered quarks/anti-quarks. Uncertainties of the flavor separated quark/anti-quark spin contributions are still relatively large~\cite{dssv}\cite{lss}.  

$W^\pm$ boson production in polarized proton+proton collisions at RHIC were proposed as a unique tool to study the spin-flavor structure of the proton at a high scale, $Q\sim M_W$~\cite{rhic_spin}, where $Q$ describes the exchanged energy. Due to the parity violation of the weak interaction, $W^{\pm}$ bosons only couple to left-handed quarks and right-handed anti-quarks. They naturally determine the helicity of the incident quarks. The charge of $W$ boson selects a specific combination of the flavor of the incoming quarks, $u+\bar{d} \rightarrow W^+$ and $d+\bar{u} \rightarrow W^-$. Subsequent leptonic decay provides a fragmentation-function-free probe of the helicity-dependent Parton Distribution Functions (PDFs). The longitudinal single-spin asymmetry is defined as $A_L=(\sigma_+-\sigma_-)/(\sigma_++\sigma_-)$, where $\sigma_{+(-)}$ is the cross section when the polarized beam has positive (negative) helicity. At leading order, the $A_L$ of $W^\pm$ are directly sensitive to $\Delta \bar{d}$ and $\Delta \bar{u}$,

\begin{equation} 
A_L^{W^+} \propto \frac{ \Delta \bar{d}(x_1) u(x_2) - \Delta u(x_1) \bar{d}(x_2) }{\bar{d}(x_1) u(x_2) - \Delta u(x_1) \bar{d}(x_2)},
\label{Eqn:ALWp}
\end{equation}
\begin{equation}
A_L^{W^-} \propto \frac{\Delta \bar{u}(x_1) d(x_2)-\Delta d(x_1) \bar{u}(x_2)}{\bar{u}(x_1)d(x_2) + d(x_1) \bar{u}(x_2)},
\label{Eqn:ALWn}
\end{equation} 
where $x_1$ and $x_2$ are the momentum fractions carried by the scattering partons. 
The $A_L^{W^+}$ ($A_L^{W^-}$) approaches $\Delta u/u$ ($\Delta d/d$) in the very forward region of $W$ rapidity, $y_W \gg 0$, and $-\Delta \bar{d}/\bar{d}$ ($-\Delta \bar{u}/\bar{u}$) in the very backward region of $W$ rapidity $y_W \ll 0$.  

First measurements of the $W$ single-spin asymmetry at RHIC were reported by the STAR\cite{PRLW} and PHENIX\cite{PhenixW} collaborations from data collected during a successful commission run at $\sqrt{s}$ = 500 GeV in 2009. In the following proton+proton running years, both the STAR\cite{PRL-2014} and PHENIX\cite{PHenixW2016} collaborations have performed further measurements of $W$ $A_L$ with increased statistics and improved beam polarization. In 2013, STAR collected an integrated luminosity of $\sim$300 pb$^{-1}$ at $\sqrt{s}$ = 510 GeV with an average beam polarization of $\sim$56\%. This is  more than three times larger than the total integrated luminosity of previous years. In this contribution, we report new preliminary results on $W$ $A_L$ from STAR data obtained in 2013.

\section{Analysis}

STAR measures the decay electrons (positrons) in $W\rightarrow e\nu$. The Time Projection Chamber (TPC) covering the full azimuth and a pseudorapidity range of $-1.3 < \eta < 1.3$, is the main tracking subsystem. It provides momenta and charge sign information for charged particles. Outside the TPC, the Barrel and Endcap Electromagnetic Calorimeters (BEMC and EEMC) covering full azimuth and pseudorapidity ranges of $-1 < \eta < 1$ and $1.1 < \eta < 2.0$ respectively, measure the energy of electrons and photons. The $W^{\pm} \rightarrow e^{\pm}\nu$ candidate event is characterized by a well isolated electron track carrying transverse energy, $E_T^e$, which exhibits the two-body decay "Jacobian Peak" near  half the $W^{\pm}$ mass. Due to the large missing energy in the opposite azimuth of the $e^\pm$ candidates, resulting from the undetected decay neutrinos, there is a significant $p_T$ imbalance when summing over all reconstructed final-state objects. In contrast, the $p_T$ vector is well balanced for  background events such as $Z/\gamma^* \rightarrow e^+e^-$ and QCD di-jet or multi-jet events. The main $W$ selection cuts used in this analysis are based on these isolation and $p_T$ imbalance features. 

\begin{figure}[t]
\centerline{\includegraphics[width=.8\textwidth]{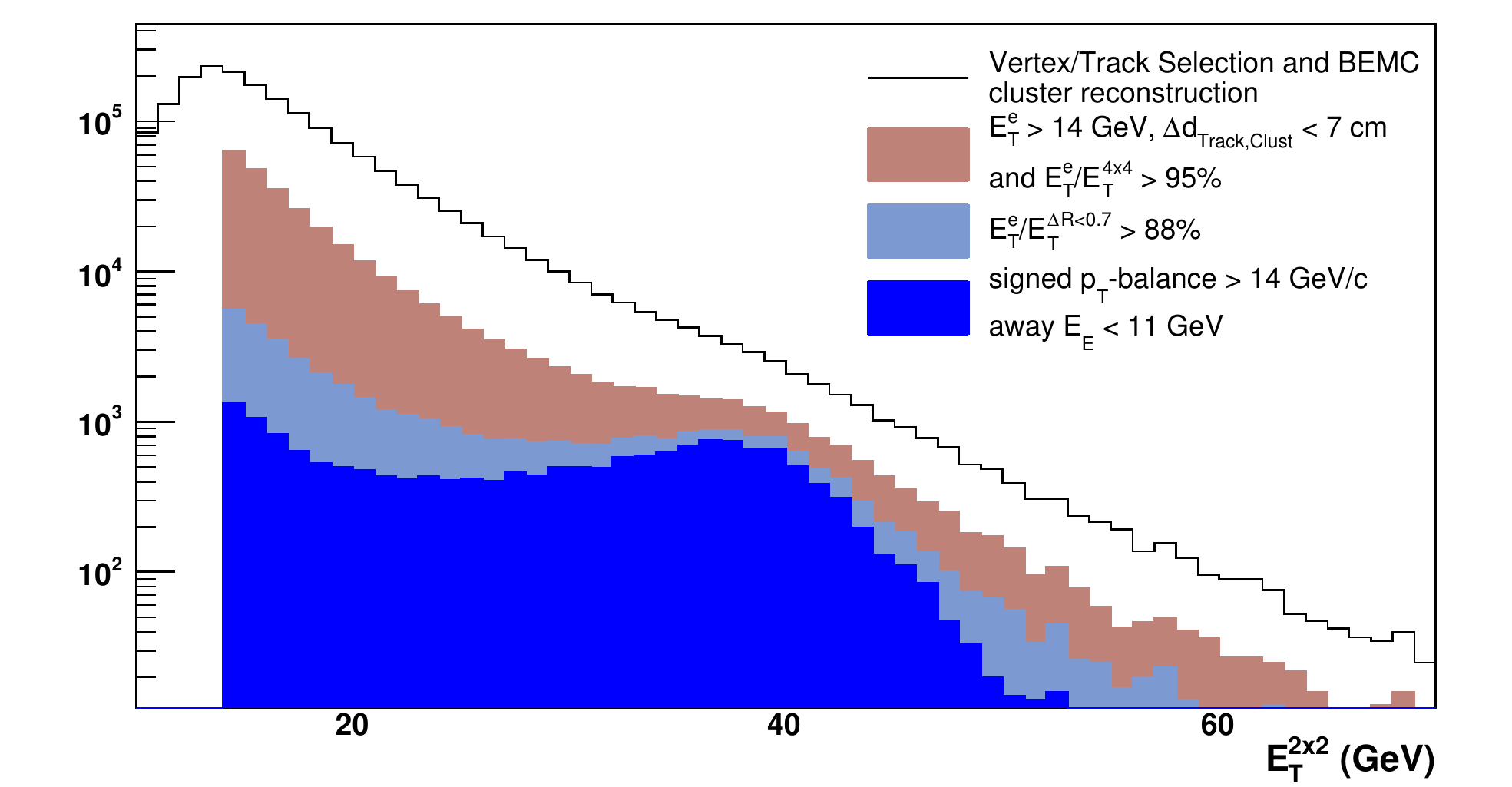}}
\caption{$E_T$ spectrum of candidate electrons at different stages of selection cuts. \label{Fig:jacobian}} 
\end{figure}

First, clusters are reconstructed from 2$\times$2 calorimeter towers pointed to by the extrapolated tracks of high transverse momentum, $p_T$ > 10 GeV. The cluster energy $E_T^{2\times2}$ is assigned to the candidate electron. Then, two stages of isolation cuts are implemented. One  requires that the 2$\times$2 cluster contains more than 95\% of the transverse energy in a 4$\times$4 cluster surrounding the 2$\times$2 cluster. Another requires the candidate electron to carry more than 88\% of the energy in the near side cone with radius $\Delta$R = 0.7, where $\Delta R = \sqrt{\Delta \phi^2 + \Delta \eta^2}$.  After that, a "signed p$_T$-balance" is constructed by projecting the $\vec{p_T^{bal}}$ to the direction of candidate electron $\vec{p_T^e}$, 
where $\vec{p_T^{bal}}$ is the vector sum of candidate electron $\vec{p_T^e}$ and the $p_T$ vectors of all reconstructed jets outside the isolation cone mentioned above. The signed p$_T$-balance is required to be larger than 14 GeV. In this analysis, the jet reconstruction used the standard anti-$k_T$ algorithm\cite{antikt}. Additionally, the total transverse energy in the azimuthally away side cone is required to be less than 11 GeV to further suppress backgrounds. Figure~\ref{Fig:jacobian} shows the candidate $E_T$ distributions at different implementation stages of selection cuts. The "Jacobian Peak" is seen to be pronounced more and more after cuts are applied. 

\begin{figure}[t]
\centerline{\includegraphics[width=.95\textwidth]{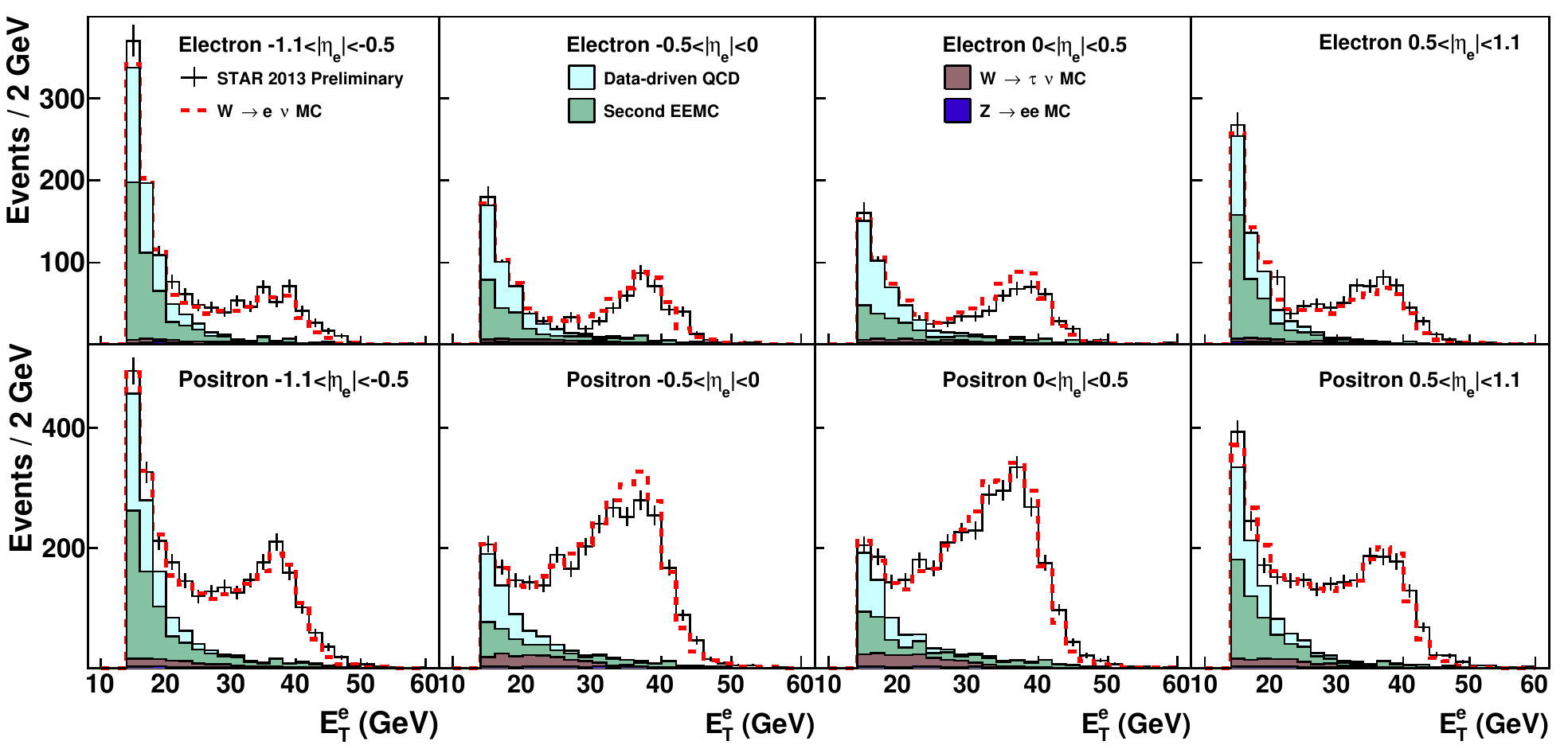}}
\vspace*{8pt}
\caption{$E_T^e$ distribution of  $W^-$ (top) and $W^+$ (bottom) candidate events (black), different background contributions with different colors(see legend), and sum of backgrounds and $W \rightarrow e\nu$ MC signal (red-dashed). \label{Fig:BEMCW}} 
\end{figure}

After all the cuts described above, there are still some residual background events under the signal "Jacobian Peak". STAR is not a 4$\pi$ coverage detector. A di-jet event or $Z/\gamma^* \rightarrow e^+e^-$ event for example, can have one of its jets or electrons outside the STAR acceptance. Such events are accepted if the detected jet or electron passes all the $W$ selection criteria . In addition, the $W$ boson can decay to $\tau+\nu$ and the $\tau$ can further decay to electrons. We do not distinguish these feed down electrons from signal electrons. To estimate the $Z/\gamma^*$ and $\tau$ contributions, Monte Carlo (MC) events were generated using the PYTHIA~\cite{pythia} generator and propagated through the STAR detector simulation framework based GEANT~\cite{geant}. Then, the sample of MC events was embedded into STAR zero-bias proton+proton events and analyzed with the same analysis algorithms. The events that passed all the $W$ selection cuts, were normalized to the integrated luminosity of the data, and were taken into account as contributions to the background. The QCD background was estimated using two procedures. The existing EEMC was used to assess the background from the corresponding uninstrumented acceptance region on the opposite side of the collision point. The remainders of the QCD background was estimated by normalizing the $E_T$ spectrum of an ideal QCD sample to the observed $E_T$ spectrum in a QCD dominated interval. In Figure~\ref{Fig:BEMCW}, the $E_T$ distributions for $W^+$ and $W^-$ are shown for different pseudorapidity intervals, where the black histograms are the raw signal and the colored histograms are for different background contributions. The $E_T$ distributions from data and simulation are  consistent. The background contribution was calculated in the signal window, 25 < $E_T$ < 50 GeV, and quantified for each pseudorapidity interval. Additional details about the analysis procedure can be found in Ref.~\cite{PRL-2014}.

\section{Results}

From the spin sorted yields of $W^{\pm}$ bosons, the longitudinal single-spin asymmetries were extracted in four pseudorapidity intervals, using
\begin{equation} 
A_L = \frac{1}{\beta}\frac{1}{P}\frac{N_+/l_+ - N_-/l_-}{N_+/l_+ + N_-/l_-}, 
\label{eq:AL}
\end{equation}
where $\beta$ quantifies the dilution due to background, $P$ is the beam polarization, $N_+(N_-)$ is the $W$ yield when the helicity of the polarized beam is positive (negative), and $l_{\pm}$ are the relative luminosity correction factors. 

\begin{figure}[t]
\centerline{\includegraphics[width=0.65\textwidth]{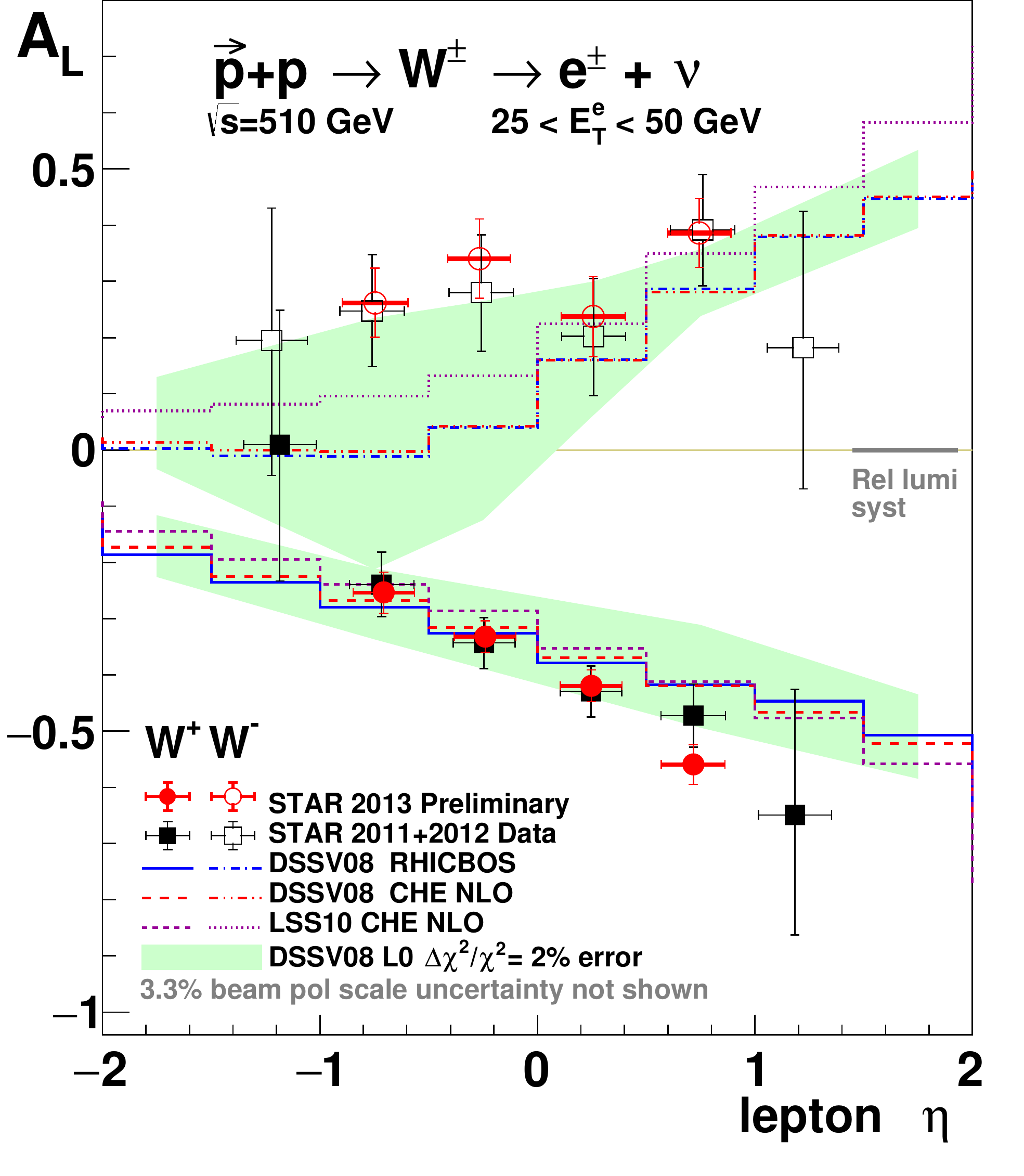}}
\vspace*{8pt}
\caption{STAR 2013 preliminary longitudinal single-spin asymmetry results for $W^\pm$ production as a function of lepton pseudorapidity, $\eta_e$, in comparison with theoretical predictions and STAR 2011+2012 results. \label{Fig:WAL} }
\end{figure}

Figure~\ref{Fig:WAL} shows STAR 2013 preliminary $W^{\pm}$ results in comparison with theoretical predictions and previous STAR results from the combined 2011 and 2012 data\cite{PRL-2014}. $A_L$ for $W^+$ is consistent with the theoretical predictions albeit decreasing slightly more steeply with increasing pseudorapidity. $A_L$ for $W^-$, however, are above the theoretical predictions at negative pseudorapidity. The 2013 preliminary results are consistent with 2011+2012 results, and have $\sim$40\% smaller uncertainties. The 2011+2012 results have been included into the global QCD analysis by NNPDF group~\cite{nnpdf}. The constraints provided by these STAR data lead to a shift in the central value of $\Delta\bar{u}$ from negative to positive in RHIC $x$ range 0.05 < $x$ < 0.2. The data favor $\Delta \bar{u} > \Delta\bar{d}$ opposite to the unpolarized distributions.  An earlier preliminary global analysis performed by the DSSV group included STAR 2012 preliminary results. They reported a similar impact on the $\Delta\bar{u}$ and $\Delta\bar{d}$ distributions~\cite{dssv+}.  STAR 2013 $W^{\pm}$ $A_L$ results have reached unprecedented precision and are expected to significantly advance our understanding of polarizations in the light quark sea. 

The author is supported partially by the MoST of China (973 program No. 2014CB845406), the Natural Science Foundation of Shandong Province, China, under Grant No.ZR2013JQ001 and the Office of Science of the U.S. Department of Energy under Contract No. DE-AC02-05CH11231. 

\end{document}